\documentclass[prc,aps,twocolumn,showpacs,amssymb,superscriptaddress,float]{revtex4}

\usepackage{amsmath}
\usepackage{graphicx}
\usepackage{dcolumn}
\usepackage{float}

\begin{document}

\title{Low Momentum Nucleon-Nucleon Interactions and Shell-Model 
Calculations}

\author{L. Coraggio}
\affiliation{Dipartimento di Scienze Fisiche, Universit\`a
di Napoli Federico II, \\ and Istituto Nazionale di Fisica Nucleare, \\
Complesso Universitario di Monte  S. Angelo, Via Cintia - I-80126 Napoli,
Italy}
\author{A. Covello}
\affiliation{Dipartimento di Scienze Fisiche, Universit\`a
di Napoli Federico II, \\ and Istituto Nazionale di Fisica Nucleare, \\
Complesso Universitario di Monte  S. Angelo, Via Cintia - I-80126 Napoli,
Italy}
\author{A. Gargano}
\affiliation{Dipartimento di Scienze Fisiche, Universit\`a
di Napoli Federico II, \\ and Istituto Nazionale di Fisica Nucleare, \\
Complesso Universitario di Monte  S. Angelo, Via Cintia - I-80126 Napoli,
Italy}
\author{N. Itaco}
\affiliation{Dipartimento di Scienze Fisiche, Universit\`a
di Napoli Federico II, \\ and Istituto Nazionale di Fisica Nucleare, \\
Complesso Universitario di Monte  S. Angelo, Via Cintia - I-80126 Napoli,
Italy}
\author{D. R. Entem}
\affiliation{Grupo de F\'{i}sica Nuclear, IUFFyM, Universidad de 
Salamanca, E-37008 Salamanca, Spain}
\author{T. T. S. Kuo}
\affiliation{Department of Physics, SUNY, Stony Brook, New York 11794}
\author{R. Machleidt}
\affiliation{Department of Physics, University of Idaho, Moscow, Idaho 83844}

\date{\today}

\begin{abstract}
In the last few years, the low-momentum nucleon-nucleon ($NN$)
interaction $V_{\rm low-k}$ derived from free-space $NN$
potentials has been successfully used in shell-model calculations.
$V_{\rm low-k}$ is a smooth potential which preserves the deuteron
binding energy  as well as the half-on-shell
$T$-matrix of the original $NN$ potential up to a momentum cutoff
$\Lambda$. 
In this paper we put to the test a new low-momentum $NN$ potential
derived from chiral perturbation theory at
next-to-next-to-next-to-leading order with a sharp low-momentum cutoff
at 2.1 fm$^{-1}$. 
Shell-model calculations for the oxygen isotopes using effective
hamiltonians derived from both types of low-momentum potential are
performed.
We find that the two potentials show the same perturbative behavior
and yield very similar results.
\end{abstract}


\maketitle

\section{Introduction}
A challenging goal of nuclear structure is to perform shell-model
calculations with single-particle (SP) energies and residual two-body 
interaction  both derived from a realistic nucleon-nucleon ($NN$)
potential $V_{NN}$.
To tackle this problem, a well-established framework is the
time-dependent degenerate linked-diagram perturbation theory as
formulated by Kuo, Lee and Ratcliff \cite{Kuo71,Kuo90}, which enables
the derivation of a shell-model effective hamiltonian $H_{\rm eff}$
starting from $V_{NN}$. 

As is well known, a main feature of $V_{NN}$ is the presence of a
built-in strong short-range repulsion dealing with high-momentum
components of the potential.
This hinders an order-by-order perturbative calculation of $H_{\rm
  eff}$ in terms of $V_{NN}$, as the matrix elements of the latter are
generally very large. 
However, in nuclear physics there is a natural separation of energy
scales, that can be used to formulate an advantageous theoretical
approach for $V_{NN}$. 
As a matter of fact, the characteristic quantum chromodynamics (QCD)
energy scale is $M_{\rm QCD} \sim 1$ GeV, while for nuclear systems
we have $M_{\rm nuc} \sim 100$ MeV \cite{vanKolck99}.

This consideration was at the origin of the seminal work of Weinberg 
\cite{Weinberg90,Weinberg91}, who introduced into nuclear physics 
the method of effective field theory (EFT) to study the $S$-matrix for
a process involving arbitrary numbers of low-momentum pions and
nucleons.
This approach is based on the known symmetries of QCD and parametrizes
the unknown dynamical details introducing a number of constants to be
determined.

Since then much work has been carried out on this subject (see for 
instance 
\cite{Lepage97,Kaplan98,Epelbaoum98,Bedaque99,vanKolck99,Haxton00}),
leading to the construction of $V_{NN}$'s based on chiral
perturbation theory that are able to reproduce accurately the $NN$ data
\cite{Entem03,Epelbaum05}. However, also these potentials cannot be
used in a perturbative nuclear structure calculation.

Inspired by EFT a new approach based on the renormalization group
(RG) has been recently introduced to derive a low-momentum
$NN$ interaction $V_{\rm low-k}$ \cite{Bogner01,Bogner02,Bogner03}. 
The starting point in the construction of $V_{\rm low-k}$ is a
realistic model for $V_{NN}$ such as the CD-Bonn \cite{Machleidt01b},
Nijmegen \cite{Stoks94}, Argonne $V18$ \cite{Wiringa95}, or N$^3$LO 
\cite{Entem03} potentials. 
A cutoff momentum $\Lambda$ that separates fast and slow modes is
then introduced and from the original $V_{NN}$ an effective potential, 
satisfying a decoupling condition between the low-and high-momentum
spaces, is derived by integrating out the high-momentum components. 
The main result is that $V_{\rm low-k}$ is a smooth potential which
preserves exactly the onshell properties of the original $V_{NN}$, and
is suitable to be used directly in nuclear structure calculations.
In the past few years, $V_{\rm low-k}$ has been fruitfully employed
in microscopic calculations within different perturbative frameworks
such as the realistic shell model 
\cite{Coraggio02b,Coraggio02c,Coraggio04,Coraggio05a,Coraggio05c,Coraggio06a}, 
the Goldstone expansion for doubly closed-shell nuclei 
\cite{Coraggio03,Coraggio05b,Coraggio06b}, and the Hartree-Fock theory
for nuclear matter calculations \cite{Kuckei03,Sedrakian03,Bogner05}.

The success of $V_{\rm low-k}$ suggests that there may be
a way to construct a low-momentum $NN$ potential which is more deeply 
rooted in EFT and can be used in a perturbative 
approach.
Namely, instead of taking the detour through a $NN$ potential with
high-momentum components, one may as well construct a low-momentum
potential from scratch using chiral perturbation theory. 
We have constructed such a potential at
next-to-next-to-next-to-leading order (N$^3$LO) using  a sharp cutoff
at 2.1 fm$^{-1}$. 
This potential reproduces the $NN$ phase shifts up to 200 MeV
laboratory energy and the deuteron binding energy. 
While $V_{\rm low-k}$ allows only for a numerical representation,
the new low-momentum potential (dubbed ${\rm N^3LOW}$) is given in
analytic form.
Moreover, the low-energy constants are explicitly known so that the
chiral three-nucleon forces consistent with ${\rm N^3LOW}$ can be
properly defined.

In order to investigate the perturbative properties of this new chiral
low-momentum potential, we perform shell-model calculations for
even oxygen isotopes.
We employ two different effective hamiltonians: one is based on the
$V_{\rm low-k}$ derived numerically from a `hard' N$^3$LO potential
\cite{Entem03} and the other on ${\rm N^3LOW}$. 

The paper is organized as follows. 
In Sec. II we give a short description of how $V_{\rm low-k}$ is
derived as well as an outline of the construction of the 
${\rm N^3LOW}$ potential.
In Sec. III a summary of the derivation of the shell-model 
$H_{\rm eff}$ is presented with some details of our calculations. 
In Sec. IV we present and discuss our results. 
Some concluding remarks are given in Sec. V.

\section{Low-momentum nucleon-nucleon potentials}

\subsection{Potential model $V_{\rm low-k}$ }

First, we outline the derivation of $V_{\rm low-k}$ 
\cite{Bogner01,Bogner02,Bogner03}.
As pointed out in the Introduction, the repulsive core contained in
$V_{NN}$ is smoothed by integrating out the high-momentum modes of 
$V_{NN}$ down to a cutoff momentum $\Lambda$. 
This integration is carried out with the requirement that the deuteron
binding energy and low-energy phase shifts of $V_{NN}$ are preserved
by $V_{\rm low-k}$. 
This is achieved  by the following $T$-matrix
equivalence approach. 
We start from the half-on-shell $T$ matrix for $V_{NN}$ 
\begin{eqnarray}
T(k',k,k^2) = V_{NN}(k',k) + ~~~~~~~~~~~~~~~~~~~~~~~~~~~~~~~\nonumber \\
~~+\mathcal{P} \int _0 ^{\infty} q^2 dq
V_{NN}(k',q) \frac{1}{k^2-q^2} T(q,k,k^2 ) ~~,
\end{eqnarray}

\noindent
where $\mathcal{P}$ denotes the principal value and  $k,~k'$, and $q$
stand for the relative momenta. 
The effective low-momentum $T$ matrix is then defined by
\begin{eqnarray}
T_{\rm low-k }(p',p,p^2) = V_{\rm low-k }(p',p) +
~~~~~~~~~~~~~~~~~~~~~~~\nonumber \\
+ \mathcal{P} \int _0
^{\Lambda} q^2 dq  V_{\rm low-k }(p',q) \frac{1}{p^2-q^2} T_{\rm 
low-k} (q,p,p^2) ~~,
\end{eqnarray}

\noindent
where the intermediate state momentum $q$ is integrated from 0 to the
momentum space cutoff $\Lambda$ and $(p',p) \leq \Lambda$. 
The above $T$ matrices are required to satisfy the condition 
\begin{equation}
T(p',p,p^2)= T_{\rm low-k }(p',p,p^2) \, ; ~~ (p',p) \leq \Lambda \,.
\end{equation}

The above equations define the effective low-momentum interaction 
$V_{\rm low-k}$, and it has been shown \cite{Bogner02} that they are
satisfied by the solution:
\begin{equation}
V_{\rm low-k} = \hat{Q} - \hat{Q}' \int \hat{Q} + \hat{Q}' \int
\hat{Q} \int \hat{Q} - \hat{Q}' \int \hat{Q} \int \hat{Q} \int \hat{Q}
+ ~...~~,
\label{vlowk}
\end{equation}

\noindent
which is the well known Kuo-Lee-Ratcliff (KLR) folded-diagram
expansion \cite{Kuo71,Kuo90}, originally designed for constructing
shell-model effective interactions.
In Eq.~(\ref{vlowk}) $\hat{Q}$ is an irreducible vertex function
whose intermediate states are all beyond $\Lambda$ and $\hat{Q}'$ is
obtained by removing from $\hat{Q}$ its terms first order in the
interaction $V_{NN}$. 
In addition to the preservation of the half-on-shell $T$ matrix, which
implies preservation of the phase shifts, this $V_{\rm low-k}$
preserves the deuteron binding energy, since eigenvalues are preserved
by the KLR effective interaction. 
For any value of $\Lambda$, the low-momentum potential of
Eq. (\ref{vlowk}) can be calculated very accurately using iteration
methods. 
Our calculation of $V_{\rm low-k}$ is performed by employing the
iteration method proposed in \cite{Andreozzi96}, which is based on the 
Lee-Suzuki similarity transformation \cite{Suzuki80}. 

The 
$V_{\rm low-k}$ 
given by the $T$-matrix equivalence approach
mentioned above is not hermitian. 
Therefore, an additional transformation is needed to make it
hermitian. 
To this end, we resort to the hermitization procedure suggested in 
\cite{Andreozzi96}, which makes use of the Cholesky decomposition of
symmetric positive definite matrices. 

\subsection{Potential model ${\rm N^3LOW}$ }
The general and fundamental reason why the $V_{\rm low-k}$ approach to
nuclear structure physics works is that the dynamics which rules
nuclear physics can be described in the framework of a low-energy
EFT. 
This nuclear EFT is characterized by the symmetries of
low-energy QCD, in particular, spontaneously broken chiral symmetry,
and the degrees of freedom relevant for nuclear physics, nucleons and
pions.
The expansion based upon this EFT has become known as chiral
perturbation theory (${\rm \chi PT}$), which is an expansion in terms
of $(Q/M_{\rm QCD})^\nu$ where $Q$ denotes the magnitude of a nucleon
three-momentum or a pion four-momentum, and $M_{\rm QCD}$ is the QCD
energy scale \cite{Weinberg90,Weinberg91}.
For this expansion to converge at a proper rate, we have to have $Q\ll
M_{\rm QCD}$. 
To enforce this, chiral $NN$ potentials are multiplied by a regulator
function that suppresses the potential for nucleon momenta $Q>\Lambda$
with $\Lambda\ll M_{\rm QCD}$.
Present chiral $NN$ potentials \cite{Entem03,Epelbaum05} typically
apply values for $\Lambda$ around 2.5 fm$^{-1}$. 

It is of course not an accident that the latter cutoff value is not
too different from what is typically used for $V_{\rm low-k}$, namely,
$\Lambda=2.1$ fm$^{-1}$.
This fact stimulates an obvious question, namely: Is it possible to
construct a chiral $NN$ potential with $\Lambda=2.1$ fm$^{-1}$?
Not surprisingly, the answer is in the affirmative.

Thus, we have constructed a $NN$ potential at N$^3$LO of chiral
perturbation theory that carries a sharp momentum cutoff at 2.1
fm$^{-1}$.
We have dubbed this potential ${\rm N^3LOW}$. 

One advantage of this potential is that it is given in analytic form. 
The analytic expressions are the same as for the ``hard'' N$^3$LO
potential by Entem and Machleidt of 2003 \cite{Entem03} and are given
in Ref. \cite{Entem02}. 
Note that the procedure that needs to be followed to construct $V_{\rm
  low-k}$ from a  free $NN$ potential (cf.\ Sec.\ II.A) can be carried
out only numerically. 

Another important issue are many-body forces. 
As demonstrated in Ref. \cite{Bogner05}, when applying a $V_{\rm
  low-k}$ in certain nuclear many-body systems, the inclusion of
three-body forces (3NF) may be crucial.
For example, nuclear matter does not saturate without a 3NF when the
two-nucleon force is represented by a low-momentum potential. 
One great advantage of $\chi$PT is that it generates nuclear two- and
many-body forces on an equal footing (for an overview of this aspect,
see Ref. \cite{Machleidt05}).
Most interaction vertices that appear in the 3NF and in the
four-nucleon force (4NF) also occur in the two-nucleon force (2NF). 
The parameters carried by these vertices are fixed in the
construction of the chiral 2NF. 
Consistency requires that for the same vertices the same parameter
values are used in the 2NF, 3NF, 4NF, \ldots .
If the 2NF is analytic, these parameters are known, and there is no
problem with their consistent proliferation to the many-body force
terms. 
However, if a potential exists only in numeric form, then those
parameters are not explicitly known and the parameters to be used in
the 3NF, 4NF, \ldots must be based upon `educated guesses'. 
The firm consistency between two- and many-body forces is lost.
Moreover, if a potential is given only in numeric form, one does not
known what order of $\chi$PT it belongs to. 
Thus, it is also not clear which orders of 3NF and 4NF to include to
be consistent with the order of the 2NF.

\begin{figure}[H]
\begin{center}
\includegraphics[scale=0.45,angle=0]{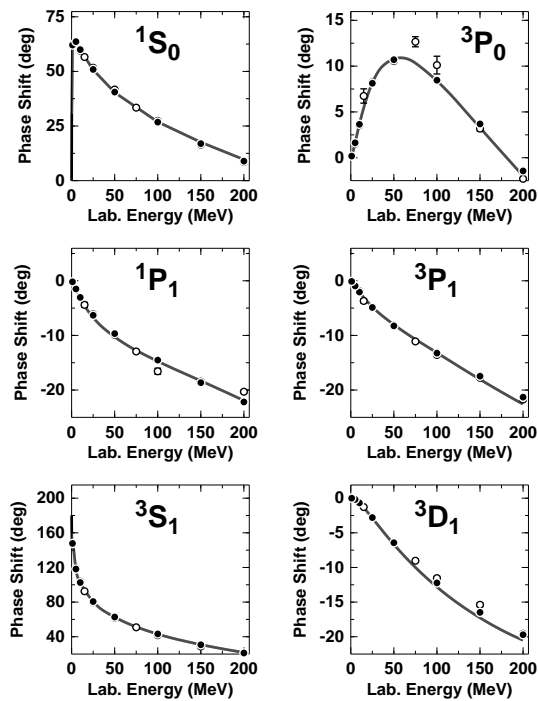}
\caption{Phase parameters of neutron-proton scattering
up to 200 MeV laboratory energy for partial waves
with total angular momentum $J\leq2$.
The solid lines show the predictions by the new low-momentum 
${\rm N^3LOW}$ potential.  
The solid dots and open circles represent the Nijmegen multienergy
$np$ phase shift analysis~\protect\cite{Nijmegen93} and the GWU/VPI
single-energy $np$ analysis SM99~\protect\cite{SM99}, respectively.} 
\label{phaseshifts1}
\end{center}
\end{figure}

\begin{figure}[H]
\begin{center}
\includegraphics[scale=0.45,angle=0]{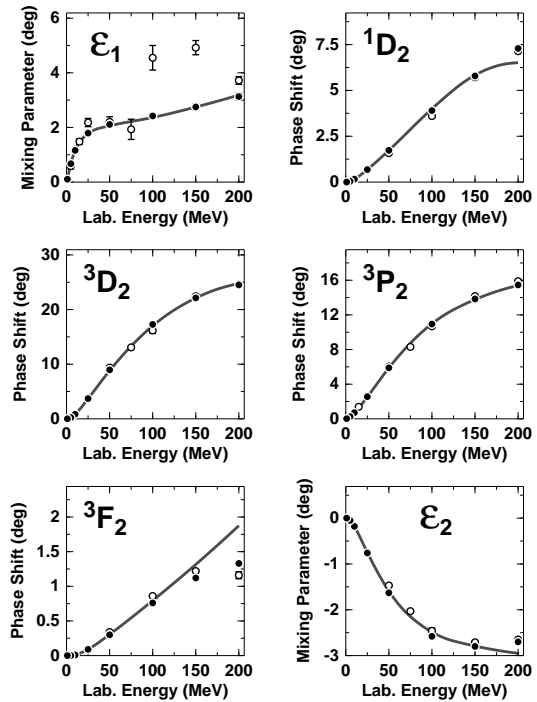}
\caption{Same as Fig. \ref{phaseshifts1}.} 
\label{phaseshifts2}
\end{center}
\end{figure}

Our newly constructed ${\rm N^3LOW}$ reproduces accurately the
empirical deuteron binding energy, the experimental low-energy
scattering parameters, and the empirical phase-shifts of $NN$
scattering up to at least 200 MeV laboratory  energy, see
Figs. \ref{phaseshifts1},\ref{phaseshifts2}. 
More details about this potential will be published
elsewhere \cite{Entem07}. 
It is the main purpose of this paper to test if the perturbative
properties of ${\rm N^3LOW}$ when applied in microscopic nuclear
structure are as good as the ones of typical $V_{\rm low-k}$ potentials.

\section{Derivation of the shell-model effective hamiltonian}
In the framework of the shell model, an auxiliary one-body potential
$U$ is introduced in order to break up the nuclear hamiltonian as the
sum of a one-body component $H_0$, which describes the independent
motion of the nucleons, and a residual interaction $H_1$:

\begin{eqnarray}
H & = & \sum_{i=1}^{A} \frac{p_i^2}{2m} + \sum_{i<j} V_{ij} = T + V = 
\nonumber \\
 ~~ & = &(T+U)+(V-U)= H_{0}+H_{1}~~.
\label{smham}
\end{eqnarray}

Once $H_0$ has been introduced, a reduced model space is defined in
terms of a finite subset of $H_0$'s eigenvectors. 
The diagonalization of the many-body hamiltonian (\ref{smham}) in an
infinite Hilbert space, that is obviously unfeasible, is then
reduced to the solution of an eigenvalue problem for an effective 
hamiltonian $H_{\rm eff}$ in a finite space.

In this paper we derive $H_{\rm eff}$ by way of the time-dependent
perturbation theory \cite{Kuo71,Kuo90}.
Namely, $H_{\rm eff}$ is expressed through the KLR folded-diagram
expansion in terms of the vertex function $\hat{Q}$-box, which is
composed of irreducible valence-linked diagrams. 
We take the $\hat{Q}$-box to be composed of one- and two-body
Goldstone diagrams through third order in $V$ \cite{Kuo81}.
Once the $\hat{Q}$-box has been calculated at this perturbative order,
the series of the folded diagrams is summed up to all orders using the
Lee-Suzuki iteration method \cite{Suzuki80}.

The hamiltonian $H_{\rm eff}$ contains one-body contributions, which
represent the effective SP energies.
In realistic shell-model calculations it is customary to use a
subtraction procedure \cite{Shurpin83} so that only the two-body terms
of $H_{\rm eff}$ are retained - the effective interaction $V_{\rm
  eff}$ - and the SP energies are taken from the experimental data.

In this work, we have followed a different approach and employed the
theoretical SP energies obtained from the calculation of $H_{\rm eff}$. 
This allows to make a consistent study of the perturbative properties
of the input potential $V$ ($V_{\rm low-k}$ or ${\rm N^3LOW}$) in the
shell-model approach.

In this regard, it is worth to point out that, owing to the presence
of the $-U$ term in $H_1$, $U$-insertion diagrams arise in the 
$\hat{Q}$-box.
In our calculation we use as auxiliary potential the harmonic
oscillator (HO), $U=\frac{1}{2} m \omega ^2 r^2 + \Delta$, and take
into account only the first-order $U$-insertion that appears in the 
collection of the one-body contributions, which is the dominant one 
\cite{Shurpin77}.

\section{Results}
We have performed shell-model calculations for even-mass oxygen
isotopes beyond the doubly-closed core $^{16}$O.
Oxygen isotopes constitute a quite interesting nuclear chain both
theoretically and experimentally, and they have been considered a
testing ground for realistic shell-model calculations since the
pioneering work by Kuo and Brown \cite{Kuo66}.
Our calculations have been carried out by using the Oslo shell-model
code \cite{EngelandSMC}

Two $H_{\rm eff}$'s have been obtained using, respectively, a 
$V_{\rm low-k}$ with a cutoff momentum $\Lambda=2.1$ fm$^{-1}$ derived
numerically from the `hard' N$^3$LO chiral $NN$ potential
\cite{Entem03}, and the new ${\rm N^3LOW}$ potential described in
Sec.\ II.B. 
The Coulomb force between proton-proton intermediate states has been
explicitly taken into account. 
For the HO parameter $\hbar \omega$ we have used the value of 14 MeV, 
as obtained from the expression  $\hbar \omega= 45 A^{-1/3} -25 
A^{-2/3}$ \cite{Blomqvist68} for $A=16$. 
The value of $\Delta$ has been chosen so that the self-energy and the
$U$-insertion diagrams almost cancel each other \cite{Shurpin77}.
The numerical value is -54 MeV yielding an unperturbed energy of the
$0s1d$ shell equal to -5 MeV, which is not far the experimental value
of the ground state (g.s.) energy of $^{17}$O relative to $^{16}$O
(-4.144 MeV \cite{Audi03}). 
However, it is worth to point out that, when including only first-order
$U$-insertions, our $H_{\rm eff}$ matrix elements do not depend on the
choice of $\Delta$.

The perturbative behavior of the input potential $V$ rules the
convergence rate of the diagrammatic series for $H_{\rm eff}$ which
has to deal with the convergence of both the order-by-order
perturbative expansion and the sum over the intermediate states in the
Goldstone diagrams. In this context, we have found it interesting to
study and compare the convergence properties of $H_{\rm eff}$  derived
from both the $V_{\rm low-k}$ and the ${\rm N^3LOW}$ potentials.

\begin{table}[H]
\caption{Ground state energies (in MeV) of $^{18}$O 
relative to $^{16}$O calculated with  $H_{\rm eff}$ derived from the
$V_{\rm low-k}$ of the ``hard'' N$^3$LO potential as a function of the
maximum number $N_{\rm max}$ of the HO quanta (see text for details). 
Results obtained at second and third order in perturbation theory are
reported.} 
\begin{ruledtabular}
\begin{tabular}{lccccccc}
\colrule
 $N_{\rm max}$ & 4 & 6 & 8 & 10 & 12 & 14 & 16 \\
2nd & -8.191 & -10.615 & -12.748 & -14.318 & -15.037 & -15.142
 & -15.168 \\
3rd & -6.617 & -8.987 & -11.344 & -13.277 & -14.294 & -14.487
 & -14.523 \\
\end{tabular}
\end{ruledtabular}
\label{vlwkconvtab1}
\end{table}

\begin{table}[H]
\caption{Same as Table \ref{vlwkconvtab1}, but with  $H_{\rm eff}$ 
derived from ${\rm N^3LOW}$.} 
\begin{ruledtabular}
\begin{tabular}{lccccccc}
\colrule
 $N_{\rm max}$ & 4 & 6 & 8 & 10 & 12 & 14 & 16 \\
2nd & -4.806 & -6.789 & -9.085 & -11.759 & -13.671 & -14.108
 & -14.162 \\
3rd & -3.547 & -5.366 & -7.683 & -10.789 & -13.339 & -13.992
 & -14.049 \\
\end{tabular}
\end{ruledtabular}
\label{n3loconvtab1}
\end{table}

In Table \ref{vlwkconvtab1} we present the $V_{\rm low-k}$ g.s. energies
of $^{18}$O relative to $^{16}$O obtained at second- and third-order 
perturbative expansion of  $H_{\rm eff}$ . In both cases the energies
are reported as a function of the maximum allowed excitation energy of
the intermediate states, expressed in terms of the oscillator quanta
$N_{\rm max}$. 
It can be seen that the g.s. energy is practically convergent at
$N_{\rm max}=12$. 
As regards the order-by-order convergence, it may  be considered quite
satisfactory, the difference between second- and third-order results
being around 4 $\%$ for $N_{\rm max}=16$, which is the largest $N_{\rm 
max}$ \ we have employed.
Similar results are obtained using the ${\rm N^3LOW}$ potential, as
shown in Table \ref{n3loconvtab1}. 
In this case for  $N_{\rm max}=16$ the difference between second- and
third-order results is less than 1 $\%$. 

\begin{table}[H]
\caption{Same as Table \ref{vlwkconvtab1}, but with  experimental SP 
energies.}
\begin{ruledtabular}
\begin{tabular}{lccccccc}
\colrule
 $N_{\rm max}$ & 4 & 6 & 8 & 10 & 12 & 14 & 16 \\
2nd & -12.201 & -12.000 & -11.847 & -11.771 & -11.750 & -11.748
 & -11.749 \\
3rd & -12.373 & -12.328 & -12.286 & -12.281 & -12.290 & -12.294
 & -12.296 \\
\end{tabular}
\end{ruledtabular}
\label{vlwkconvtab2}
\end{table}

\begin{table}[H]
\caption{Same as Table \ref{n3loconvtab1}, but with experimental SP 
energies.}
\begin{ruledtabular}
\begin{tabular}{lccccccc}
\colrule
 $N_{\rm max}$ & 4 & 6 & 8 & 10 & 12 & 14 & 16 \\
2nd & -12.386 & -12.202 & -12.072 & -12.002 & -11.973 & -11.968
 & -11.967 \\
3rd & -12.663 & -12.637 & -12.612 & -12.656 & -12.722 & -12.737
 & -12.739 \\
\end{tabular}
\end{ruledtabular}
\label{n3loconvtab2}
\end{table}

It is now worth to comment on the dimension of the intermediate-state space.
We have found  that, when increasing $N_{\rm max}$, the matrix
elements of the two-body effective interaction $V_{\rm eff}$ (TBME)
are much more stable than the calculated  SP energies. 
In this connection, an inspection of Tables \ref{vlwkconvtab2}
and \ref{n3loconvtab2}, where we present results obtained 
using experimental SP energies, evidences the rapid convergence of the 
g.s. energy with $N_{\rm max}$. 
This supports the choice of moderately large intermediate state
spaces in standard realistic shell-model calculations, where in general
experimental SP energies are employed.

A comparison between the results obtained from the $V_{\rm low-k}$ and
the ${\rm N^3LOW}$ potentials (see Tables \ref{vlwkconvtab1} and
\ref{n3loconvtab1}, respectively) shows that at third order, with a
sufficiently large number of intermediate states, the two
interactions predict very close values for the relative binding energy
of $^{18}$O. 
Note that the experimental value is $-12.187$ MeV \cite{Audi03}.
From now on we refer to calculations with $N_{\rm max}=16$ and
$\hat{Q}$-box diagrams up to third order in perturbation theory.

\begin{table}[H]
\caption{$J^{\pi}=0^+$ TBME of $H_{\rm eff}$ (in MeV) obtained from the
$V_{\rm low-k}$ of the ``hard'' N$^3$LO potential and ${\rm N^3LOW}$ 
at third order in perturbation theory with $N_{\rm max}=16$. 
They are compared with the USDA TBME of Brown and Richter 
\cite{Richter06}.}
\begin{ruledtabular}
\begin{tabular}{cccc}
\colrule
 configuration & $V_{\rm low-k}$ & ${\rm N^3LOW}$ & USDA \\
 $ (0d5/2)^2~(0d5/2)^2$  & -2.435 & -2.689 & -2.480 \\
 $ (0d5/2)^2~(0d3/2)^2$  & -3.464 & -3.741 & -3.569 \\
 $ (0d5/2)^2~(1s1/2)^2$  & -1.269 & -1.337 & -1.157 \\
 $ (0d3/2)^2~(0d3/2)^2$  & -0.845 & -1.147 & -1.505 \\
 $ (0d3/2)^2~(1s1/2)^2$  & -0.862 & -0.967 & -0.983 \\
 $ (1s1/2)^2~(1s1/2)^2$  & -2.385 & -2.563 & -1.846 \\
\end{tabular}
\end{ruledtabular}
\label{mtxtab}
\end{table}

\begin{table}[H]
\caption{Calculated SP relative energies of $H_{\rm eff}$ (in MeV)   
obtained from the $V_{\rm low-k}$ of the ``hard'' N$^3$LO potential
and ${\rm N^3LOW}$ at third order in perturbation theory with $N_{\rm
  max}=16$. 
They are compared with the USDA SP energies of Brown and Richter
\cite{Richter06} and the experimental ones.
The values in parenthesis are the absolute SP energies.} 
\begin{ruledtabular}
\begin{tabular}{lcccc}
\colrule
 orbital & $V_{\rm low-k}$ & ${\rm N^3LOW}$ & USDA & Expt \\
 $\nu 0d5/2$  & 0.0 (-5.425) & 0.0 (-4.909) & 0.0 (-3.944) & 0.0 (-4.144) \\
 $\nu 0d3/2$  & 7.323        & 7.117        & 5.924        & 5.085  \\
 $\nu 1s1/2$  & 1.257        & 0.818        & 0.883        & 0.871  \\
 \end{tabular}
\end{ruledtabular}
\label{spetab}
\end{table}

In Table \ref{mtxtab} and \ref{spetab}, we present the $J=0^+$ TBME
and the SP energies obtained from the $V_{\rm low-k}$ and the ${\rm
  N^3LOW}$ potentials. 
It turns out  that the results from  the two potentials are quite
similar, the largest difference regarding the absolute energy  of the
$d_{5/2}$ SP level. 
In the last column of Table \ref{mtxtab} and \ref{spetab} we also show 
the $J^{\pi}=0^{+}$ TBME and SP energies obtained by Brown and Richter
\cite{Richter06} from a least-squares fit of a large set of energy
data for the $sd$-shell nuclei. 
The reported values refer to the USDA hamiltonian, see
Ref. \cite{Richter06} for details. 
It is interesting to note that the TBME of USDA are closer to ours
than those of the original USD \cite{Wildenthal84,Brown88}, which was
based on a smaller set of data. 
Regarding the SP energies, we see that the USDA absolute energy of
the $d_{5/2}$ state is in good agreement with  the experimental
g.s. state energy of $^{17}$O, while both our calculations give about
one MeV more binding. 
On the other hand, our calculated and the USDA excitation energies of
the $s_{1/2}$ state come close to each other and do not differ
significantly from the experimental value.
As regards the relative energy of the $d_{3/2}$ state, the USDA value
is larger than the experimental one, in line with our findings. 

\begin{figure}[H]
\begin{center}
\includegraphics[scale=0.45,angle=0]{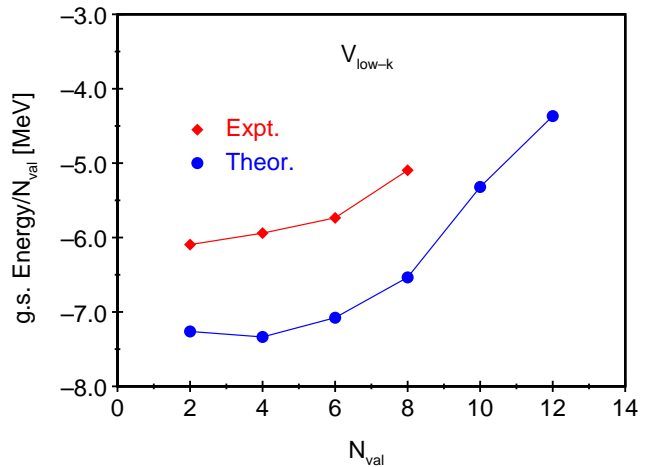}
\caption{(Color online)  Experimental and calculated ground-state 
energy per valence neutron for even oxygen isotopes from $A=18$ to
28. 
$N_{\rm val}$ is the number of valence neutrons. 
The calculated values are obtained with $H_{\rm eff}$ derived from the
$V_{\rm low-k}$ of the ``hard'' N$^3$LO potential at third order in 
perturbation theory, with $N_{\rm max}=16$.}
\label{gsvlwk}
\end{center}
\end{figure}

\begin{figure}[H]
\begin{center}
\includegraphics[scale=0.45,angle=0]{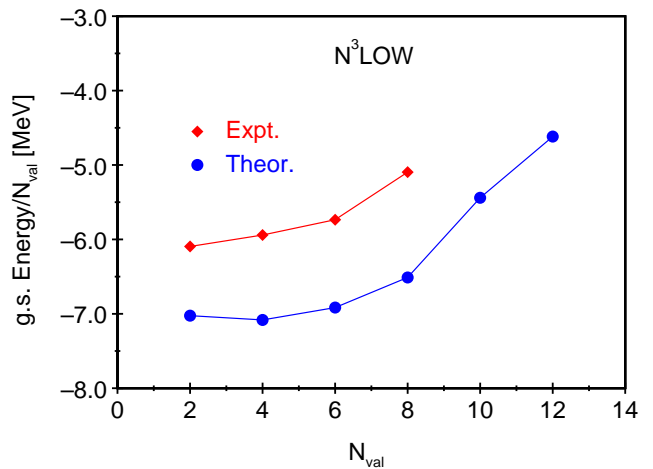}
\caption{(Color online) Same as Fig. \ref{gsvlwk}, but  with $H_{\rm 
eff}$ derived from ${\rm N^3LOW}$.}
\label{gsn3lo}
\end{center}
\end{figure}

\begin{figure}[H]
\begin{center}
\includegraphics[scale=0.45,angle=0]{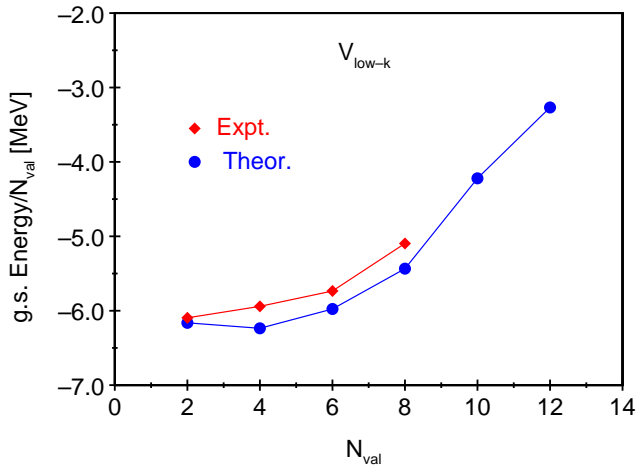}
\caption{(Color online)  Same as Fig. \ref{gsvlwk}, but with SP
energies shifted upward by 1.1 MeV.}
\label{gsvlwkshift}
\end{center}
\end{figure}

\begin{figure}[H]
\begin{center}
\includegraphics[scale=0.45,angle=0]{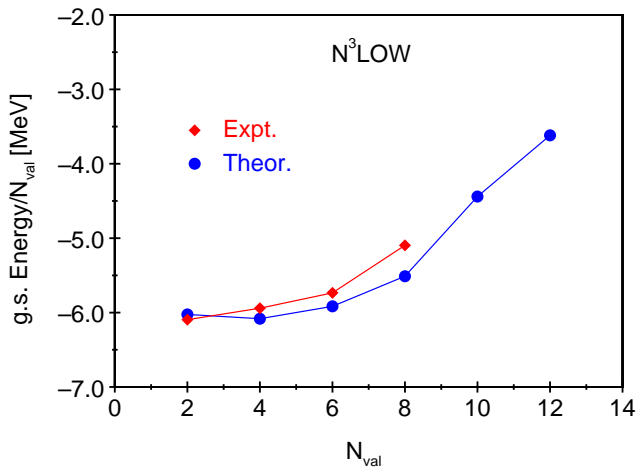}
\caption{(Color online)  Same as Fig. \ref{gsn3lo}, but with SP
energies shifted upward by 1 MeV.}
\label{gsn3loshift}
\end{center}
\end{figure}

It is worth recalling that $H_{\rm eff}$, which has been derived in
the framework of the KLR approach, is defined for the two
valence-nucleon problem. 
It is of interest, however, to test the theoretical SP energies, as
given in Table \ref{spetab}, and the TBME in systems with many
valence nucleons. 
To this end, we have calculated for even oxygen isotopes the
g.s. energies relative to $^{16}$O core up to $A$=28, and the
excitation energies of the first $2^+$ state up to $A$=24. 
In Figs. \ref{gsvlwk} and \ref{gsn3lo} the experimental g.s. energies
per valence nucleon are compared with the calculated ones for the
$V_{\rm low-k}$ and the ${\rm N^3LOW}$ potential, respectively. 
The predictions from both potentials yield binding energies which are
larger than the experimental ones, with the difference remaining
almost constant when increasing the number of valence neutrons $N_{\rm 
val}$.
In fact, if we shift the $V_{\rm low-k}$ and ${\rm N^3LOW}$ SP spectra
upward by 1.1 MeV and 1 MeV, respectively, such as to reproduce the
experimental $^{18}$O binding energy, then we obtain a very good
agreement, as shown in Figs. \ref{gsvlwkshift} and  \ref{gsn3loshift}.

This result is very interesting: our calculations fail to predict that
$^{26}$O and $^{28}$O are unbound to 2-neutron decay as indicated by
experimental studies \cite{Fauerbach96,Tarasov97}. 
However, when the SP energies are shifted upward, then we obtain the
correct binding properties, as it may be seen from
Figs. \ref{gsvlwkshift} and  \ref{gsn3loshift}. 
In some recent papers \cite{Brown01,Ellis05} it has been argued that
with a realistic shell-model interaction it is not possible to
reproduce the correct behavior of the binding energy of oxygen
isotopes. 
In Refs. \cite{Ellis05,Zuker03} this has been ascribed to the
lack of both genuine and effective three-body correlations.
Our results suggest that a major factor is the lack of
a real three-body force.
In fact, as discussed above, we have obtained a good description of
the binding energy of oxygen isotopes by modifying the  SP spectrum so
as to reproduce the  g.s. energy  of two-valence nucleon $^{18}$O.
The last quantity is obviously not affected by effective three-body
correlations, but it is sensitive to genuine three-body forces.

\begin{figure}[H]
\begin{center}
\includegraphics[scale=0.45,angle=0]{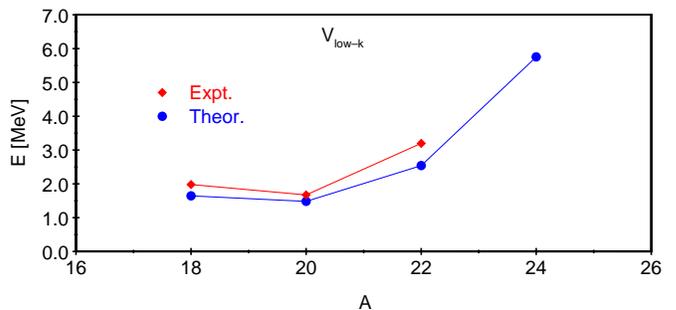}
\caption{(Color online) Experimental and calculated excitation energy 
of the first $2^+$ state  for oxygen isotopes. 
The calculated values are obtained with  $H_{\rm eff}$ derived from the
$V_{\rm low-k}$ of the ``hard'' N$^3$LO potential. Experimental data
are taken from \cite{nndc}.}
\label{2pvlwk}
\end{center}
\end{figure}

\begin{figure}[H]
\begin{center}
\includegraphics[scale=0.45,angle=0]{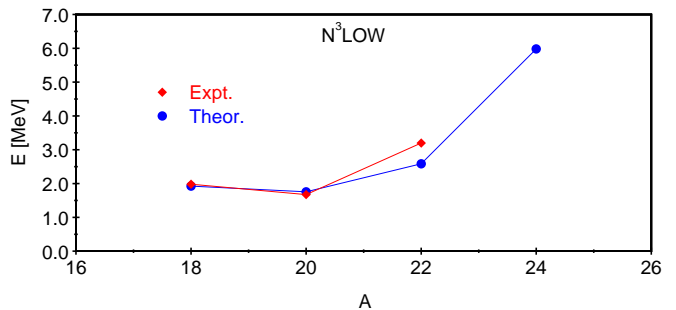}
\caption{(Color online) Same as in Fig. \ref{2pvlwk}, but with
  $H_{\rm eff}$ derived from ${\rm N^3LOW}$.} 
\label{2pn3lo}
\end{center}
\end{figure}

In Figs. \ref{2pvlwk} and  \ref{2pn3lo} we compare the experimental
and calculated excitation energies of the first $2^+$ states for the
$V_{\rm low-k}$ and the ${\rm N^3LOW}$ potential, respectively.
In both cases the experimental behavior as a function of $A$ is well
reproduced. 

\section{Summary and conclusions}
In this paper, we have performed shell-model calculations employing 
an effective hamiltonian obtained from a new low-momentum potential,
dubbed ${\rm N^3LOW}$, which is derived in the framework of chiral
perturbation theory at next-to-next-to-next-to-leading order with a
sharp cutoff at 2.1 fm$^{-1}$.

We have studied the convergence properties of this potential by
calculating the ground-state energy of $^{18}$O, and have compared the
results with those obtained using the $V_{\rm low-k}$ derived from the
``hard'' N$^{3}$LO potential of Entem and Machleidt \cite{Entem03}. 
It has turned out that the two low-momentum potentials show the same
perturbative behavior. 
We have also performed calculations for the heavier oxygen isotopes,
obtaining binding energies which are very similar for the two
potentials, but larger than the experimental ones. 
However, we have found out that by introducing an empirical shift of
the SP spectrum the ground-state energies are well reproduced up to
$^{24}$O, while $^{26}$O and   $^{28}$O are unbound to two-neutron
decay, in agreement with the experimental findings.

The main purpose of this work was to test the new chiral
low-momentum $NN$ potential N$^{3}$LOW. 
Unlike $V_{\rm low-k}$, this potential has the desirable feature that
it is given in analytic form. 
We have shown here that it is suitable to be applied perturbatively in
microscopic nuclear structure calculations and yields results which
come quite close to those obtained from the $V_{\rm low-k}$ derived
from the ``hard'' N$^{3}$LO potential. 

After this first successful application, there is now motivation to
pursue further nuclear structure calculations with this new
low-momentum potential. 

\begin{acknowledgments}
This work was supported in part by the Italian Ministero
dell'Istruzione, dell'Universit\`a e della Ricerca  (MIUR), by the
U.S. DOE Grant No. DE-FG02-88ER40388, by the U.S. NSF Grant 
No. PHY-0099444, by the Ministerio de Ciencia y Tecnonolog\'\i a under
Contract No. FPA2004-05616, and by the Junta de Castilla y Le\'on
under Contract No.~SA-104/04. 
\end{acknowledgments}

\bibliographystyle{apsrev}
\bibliography{biblio}

\begin{thebibliography}{53}
\expandafter\ifx\csname natexlab\endcsname\relax\def\natexlab#1{#1}\fi
\expandafter\ifx\csname bibnamefont\endcsname\relax
  \def\bibnamefont#1{#1}\fi
\expandafter\ifx\csname bibfnamefont\endcsname\relax
  \def\bibfnamefont#1{#1}\fi
\expandafter\ifx\csname citenamefont\endcsname\relax
  \def\citenamefont#1{#1}\fi
\expandafter\ifx\csname url\endcsname\relax
  \def\url#1{\texttt{#1}}\fi
\expandafter\ifx\csname urlprefix\endcsname\relax\def\urlprefix{URL }\fi
\providecommand{\bibinfo}[2]{#2}
\providecommand{\eprint}[2][]{\url{#2}}

\bibitem[{\citenamefont{Kuo et~al.}(1971)\citenamefont{Kuo, Lee, and
  Ratcliff}}]{Kuo71}
\bibinfo{author}{\bibfnamefont{T.~T.~S.} \bibnamefont{Kuo}},
  \bibinfo{author}{\bibfnamefont{S.~Y.} \bibnamefont{Lee}}, \bibnamefont{and}
  \bibinfo{author}{\bibfnamefont{K.~F.} \bibnamefont{Ratcliff}},
  \bibinfo{journal}{Nucl. Phys. A} \textbf{\bibinfo{volume}{176}},
  \bibinfo{pages}{65} (\bibinfo{year}{1971}).

\bibitem[{\citenamefont{Kuo and Osnes}(1990)}]{Kuo90}
\bibinfo{author}{\bibfnamefont{T.~T.~S.} \bibnamefont{Kuo}} \bibnamefont{and}
  \bibinfo{author}{\bibfnamefont{E.}~\bibnamefont{Osnes}},
  \emph{\bibinfo{title}{Lecture Notes in Physics, vol. 364}}
  (\bibinfo{publisher}{Springer-Verlag, Berlin}, \bibinfo{year}{1990}).

\bibitem[{\citenamefont{van Kolck}(1999)}]{vanKolck99}
\bibinfo{author}{\bibfnamefont{U.}~\bibnamefont{van Kolck}},
  \bibinfo{journal}{Prog. Part. Nucl. Phys.} \textbf{\bibinfo{volume}{43}},
  \bibinfo{pages}{337} (\bibinfo{year}{1999}).

\bibitem[{\citenamefont{Weinberg}(1990)}]{Weinberg90}
\bibinfo{author}{\bibfnamefont{S.}~\bibnamefont{Weinberg}},
  \bibinfo{journal}{Phys. Lett. B} \textbf{\bibinfo{volume}{251}},
  \bibinfo{pages}{288} (\bibinfo{year}{1990}).

\bibitem[{\citenamefont{Weinberg}(1991)}]{Weinberg91}
\bibinfo{author}{\bibfnamefont{S.}~\bibnamefont{Weinberg}},
  \bibinfo{journal}{Nucl. Phys. B} \textbf{\bibinfo{volume}{363}},
  \bibinfo{pages}{3} (\bibinfo{year}{1991}).

\bibitem[{\citenamefont{Lepage}(1997)}]{Lepage97}
\bibinfo{author}{\bibfnamefont{G.~P.} \bibnamefont{Lepage}}, in
  \emph{\bibinfo{booktitle}{Nuclear Physics: Proceedings of the VIII Jorge
  Andre' Swieca Summer School}}, edited by
  \bibinfo{editor}{\bibfnamefont{C.~A.} \bibnamefont{Bertulani}},
  \bibinfo{editor}{\bibfnamefont{M.~E.} \bibnamefont{Bracco}},
  \bibinfo{editor}{\bibfnamefont{B.~V.} \bibnamefont{Carlson}},
  \bibnamefont{and} \bibinfo{editor}{\bibfnamefont{M.}~\bibnamefont{Nielsen}}
  (\bibinfo{publisher}{World Scientific, Singapore}, \bibinfo{year}{1997}), p.
  \bibinfo{pages}{135}.

\bibitem[{\citenamefont{Kaplan et~al.}(1998)\citenamefont{Kaplan, Savage, and
  Wise}}]{Kaplan98}
\bibinfo{author}{\bibfnamefont{D.~B.} \bibnamefont{Kaplan}},
  \bibinfo{author}{\bibfnamefont{M.~J.} \bibnamefont{Savage}},
  \bibnamefont{and} \bibinfo{author}{\bibfnamefont{M.~B.} \bibnamefont{Wise}},
  \bibinfo{journal}{Phys. Lett. B} \textbf{\bibinfo{volume}{424}},
  \bibinfo{pages}{390} (\bibinfo{year}{1998}).

\bibitem[{\citenamefont{Epelbaoum et~al.}(1998)\citenamefont{Epelbaoum,
  Gl{\"o}ckle, and Meissner}}]{Epelbaoum98}
\bibinfo{author}{\bibfnamefont{E.}~\bibnamefont{Epelbaoum}},
  \bibinfo{author}{\bibfnamefont{W.}~\bibnamefont{Gl{\"o}ckle}},
  \bibnamefont{and} \bibinfo{author}{\bibfnamefont{U.-G.}
  \bibnamefont{Meissner}}, \bibinfo{journal}{Nucl. Phys. A}
  \textbf{\bibinfo{volume}{637}}, \bibinfo{pages}{107} (\bibinfo{year}{1998}).

\bibitem[{\citenamefont{Bedaque et~al.}(1999)\citenamefont{Bedaque, Savage,
  Seki, and van Kolck}}]{Bedaque99}
\bibinfo{editor}{\bibfnamefont{P.~F.} \bibnamefont{Bedaque}},
  \bibinfo{editor}{\bibfnamefont{M.~J.} \bibnamefont{Savage}},
  \bibinfo{editor}{\bibfnamefont{R.}~\bibnamefont{Seki}}, \bibnamefont{and}
  \bibinfo{editor}{\bibfnamefont{U.}~\bibnamefont{van Kolck}}, eds.,
  \emph{\bibinfo{title}{Nuclear Physics with Effective Field Theory II}},
  vol.~\bibinfo{volume}{9} of \emph{\bibinfo{series}{Proceedings from Institute
  for Nuclear Theory}} (\bibinfo{publisher}{World Scientific, Singapore},
  \bibinfo{year}{1999}).

\bibitem[{\citenamefont{Haxton and Song}(2000)}]{Haxton00}
\bibinfo{author}{\bibfnamefont{W.~C.} \bibnamefont{Haxton}} \bibnamefont{and}
  \bibinfo{author}{\bibfnamefont{C.~L.} \bibnamefont{Song}},
  \bibinfo{journal}{Phys. Rev. Lett.} \textbf{\bibinfo{volume}{84}},
  \bibinfo{pages}{5484} (\bibinfo{year}{2000}).

\bibitem[{\citenamefont{Entem and Machleidt}(2003)}]{Entem03}
\bibinfo{author}{\bibfnamefont{D.~R.} \bibnamefont{Entem}} \bibnamefont{and}
  \bibinfo{author}{\bibfnamefont{R.}~\bibnamefont{Machleidt}},
  \bibinfo{journal}{Phys. Rev. C} \textbf{\bibinfo{volume}{68}},
  \bibinfo{pages}{041001(R)} (\bibinfo{year}{2003}).

\bibitem[{\citenamefont{Epelbaum et~al.}(2005)\citenamefont{Epelbaum,
  Gl{\"o}ckle, and Meissner}}]{Epelbaum05}
\bibinfo{author}{\bibfnamefont{E.}~\bibnamefont{Epelbaum}},
  \bibinfo{author}{\bibfnamefont{W.}~\bibnamefont{Gl{\"o}ckle}},
  \bibnamefont{and} \bibinfo{author}{\bibfnamefont{U.-G.}
  \bibnamefont{Meissner}}, \bibinfo{journal}{Nucl. Phys. A}
  \textbf{\bibinfo{volume}{747}}, \bibinfo{pages}{362} (\bibinfo{year}{2005}).

\bibitem[{\citenamefont{Bogner et~al.}(2001)\citenamefont{Bogner, Kuo, and
  Coraggio}}]{Bogner01}
\bibinfo{author}{\bibfnamefont{S.}~\bibnamefont{Bogner}},
  \bibinfo{author}{\bibfnamefont{T.~T.~S.} \bibnamefont{Kuo}},
  \bibnamefont{and} \bibinfo{author}{\bibfnamefont{L.}~\bibnamefont{Coraggio}},
  \bibinfo{journal}{Nucl. Phys. A} \textbf{\bibinfo{volume}{684}},
  \bibinfo{pages}{432c} (\bibinfo{year}{2001}).

\bibitem[{\citenamefont{Bogner et~al.}(2002)\citenamefont{Bogner, Kuo,
  Coraggio, Covello, and Itaco}}]{Bogner02}
\bibinfo{author}{\bibfnamefont{S.}~\bibnamefont{Bogner}},
  \bibinfo{author}{\bibfnamefont{T.~T.~S.} \bibnamefont{Kuo}},
  \bibinfo{author}{\bibfnamefont{L.}~\bibnamefont{Coraggio}},
  \bibinfo{author}{\bibfnamefont{A.}~\bibnamefont{Covello}}, \bibnamefont{and}
  \bibinfo{author}{\bibfnamefont{N.}~\bibnamefont{Itaco}},
  \bibinfo{journal}{Phys. Rev. C} \textbf{\bibinfo{volume}{65}},
  \bibinfo{pages}{051301(R)} (\bibinfo{year}{2002}).

\bibitem[{\citenamefont{Bogner et~al.}(2003)\citenamefont{Bogner, Kuo, and
  Schwenk}}]{Bogner03}
\bibinfo{author}{\bibfnamefont{S.}~\bibnamefont{Bogner}},
  \bibinfo{author}{\bibfnamefont{T.~T.~S.} \bibnamefont{Kuo}},
  \bibnamefont{and} \bibinfo{author}{\bibfnamefont{A.}~\bibnamefont{Schwenk}},
  \bibinfo{journal}{Phys. Rep.} \textbf{\bibinfo{volume}{386}},
  \bibinfo{pages}{1} (\bibinfo{year}{2003}).

\bibitem[{\citenamefont{Machleidt}(2001)}]{Machleidt01b}
\bibinfo{author}{\bibfnamefont{R.}~\bibnamefont{Machleidt}},
  \bibinfo{journal}{Phys. Rev. C} \textbf{\bibinfo{volume}{63}},
  \bibinfo{pages}{024001} (\bibinfo{year}{2001}).

\bibitem[{\citenamefont{Stoks et~al.}(1994)\citenamefont{Stoks, Klomp,
  Terheggen, and Swart}}]{Stoks94}
\bibinfo{author}{\bibfnamefont{V.~G.~J.} \bibnamefont{Stoks}},
  \bibinfo{author}{\bibfnamefont{R.~A.~M.} \bibnamefont{Klomp}},
  \bibinfo{author}{\bibfnamefont{C.~P.~F.} \bibnamefont{Terheggen}},
  \bibnamefont{and} \bibinfo{author}{\bibfnamefont{J.~J.~D.}
  \bibnamefont{Swart}}, \bibinfo{journal}{Phys. Rev. C}
  \textbf{\bibinfo{volume}{49}}, \bibinfo{pages}{2950} (\bibinfo{year}{1994}).

\bibitem[{\citenamefont{Wiringa et~al.}(1995)\citenamefont{Wiringa, Stoks, and
  Schiavilla}}]{Wiringa95}
\bibinfo{author}{\bibfnamefont{R.~B.} \bibnamefont{Wiringa}},
  \bibinfo{author}{\bibfnamefont{V.~G.~J.} \bibnamefont{Stoks}},
  \bibnamefont{and}
  \bibinfo{author}{\bibfnamefont{R.}~\bibnamefont{Schiavilla}},
  \bibinfo{journal}{Phys. Rev. C} \textbf{\bibinfo{volume}{51}},
  \bibinfo{pages}{38} (\bibinfo{year}{1995}).

\bibitem[{\citenamefont{Coraggio
  et~al.}(2002{\natexlab{a}})\citenamefont{Coraggio, Covello, Gargano, and
  Itaco}}]{Coraggio02b}
\bibinfo{author}{\bibfnamefont{L.}~\bibnamefont{Coraggio}},
  \bibinfo{author}{\bibfnamefont{A.}~\bibnamefont{Covello}},
  \bibinfo{author}{\bibfnamefont{A.}~\bibnamefont{Gargano}}, \bibnamefont{and}
  \bibinfo{author}{\bibfnamefont{N.}~\bibnamefont{Itaco}},
  \bibinfo{journal}{Phys. Rev. C} \textbf{\bibinfo{volume}{66}},
  \bibinfo{pages}{064311} (\bibinfo{year}{2002}{\natexlab{a}}).

\bibitem[{\citenamefont{Coraggio
  et~al.}(2002{\natexlab{b}})\citenamefont{Coraggio, Covello, Gargano, Itaco,
  Kuo, Entem, and Machleidt}}]{Coraggio02c}
\bibinfo{author}{\bibfnamefont{L.}~\bibnamefont{Coraggio}},
  \bibinfo{author}{\bibfnamefont{A.}~\bibnamefont{Covello}},
  \bibinfo{author}{\bibfnamefont{A.}~\bibnamefont{Gargano}},
  \bibinfo{author}{\bibfnamefont{N.}~\bibnamefont{Itaco}},
  \bibinfo{author}{\bibfnamefont{T.~T.~S.} \bibnamefont{Kuo}},
  \bibinfo{author}{\bibfnamefont{D.~R.} \bibnamefont{Entem}}, \bibnamefont{and}
  \bibinfo{author}{\bibfnamefont{R.}~\bibnamefont{Machleidt}},
  \bibinfo{journal}{Phys. Rev. C} \textbf{\bibinfo{volume}{66}},
  \bibinfo{pages}{021303(R)} (\bibinfo{year}{2002}{\natexlab{b}}).

\bibitem[{\citenamefont{Coraggio et~al.}(2004)\citenamefont{Coraggio, Covello,
  Gargano, and Itaco}}]{Coraggio04}
\bibinfo{author}{\bibfnamefont{L.}~\bibnamefont{Coraggio}},
  \bibinfo{author}{\bibfnamefont{A.}~\bibnamefont{Covello}},
  \bibinfo{author}{\bibfnamefont{A.}~\bibnamefont{Gargano}}, \bibnamefont{and}
  \bibinfo{author}{\bibfnamefont{N.}~\bibnamefont{Itaco}},
  \bibinfo{journal}{Phys. Rev. C} \textbf{\bibinfo{volume}{70}},
  \bibinfo{pages}{034310} (\bibinfo{year}{2004}).

\bibitem[{\citenamefont{Coraggio
  et~al.}(2005{\natexlab{a}})\citenamefont{Coraggio, Covello, Gargano, and
  Itaco}}]{Coraggio05a}
\bibinfo{author}{\bibfnamefont{L.}~\bibnamefont{Coraggio}},
  \bibinfo{author}{\bibfnamefont{A.}~\bibnamefont{Covello}},
  \bibinfo{author}{\bibfnamefont{A.}~\bibnamefont{Gargano}}, \bibnamefont{and}
  \bibinfo{author}{\bibfnamefont{N.}~\bibnamefont{Itaco}},
  \bibinfo{journal}{Phys. Rev. C} \textbf{\bibinfo{volume}{72}},
  \bibinfo{pages}{057302} (\bibinfo{year}{2005}{\natexlab{a}}).

\bibitem[{\citenamefont{Coraggio and Itaco}(2005)}]{Coraggio05c}
\bibinfo{author}{\bibfnamefont{L.}~\bibnamefont{Coraggio}} \bibnamefont{and}
  \bibinfo{author}{\bibfnamefont{N.}~\bibnamefont{Itaco}},
  \bibinfo{journal}{Phys. Lett. B} \textbf{\bibinfo{volume}{616}},
  \bibinfo{pages}{43} (\bibinfo{year}{2005}).

\bibitem[{\citenamefont{Coraggio
  et~al.}(2006{\natexlab{a}})\citenamefont{Coraggio, Covello, Gargano, and
  Itaco}}]{Coraggio06a}
\bibinfo{author}{\bibfnamefont{L.}~\bibnamefont{Coraggio}},
  \bibinfo{author}{\bibfnamefont{A.}~\bibnamefont{Covello}},
  \bibinfo{author}{\bibfnamefont{A.}~\bibnamefont{Gargano}}, \bibnamefont{and}
  \bibinfo{author}{\bibfnamefont{N.}~\bibnamefont{Itaco}},
  \bibinfo{journal}{Phys. Rev. C} \textbf{\bibinfo{volume}{73}},
  \bibinfo{pages}{031302(R)} (\bibinfo{year}{2006}{\natexlab{a}}).

\bibitem[{\citenamefont{Coraggio et~al.}(2003)\citenamefont{Coraggio, Itaco,
  Covello, Gargano, and Kuo}}]{Coraggio03}
\bibinfo{author}{\bibfnamefont{L.}~\bibnamefont{Coraggio}},
  \bibinfo{author}{\bibfnamefont{N.}~\bibnamefont{Itaco}},
  \bibinfo{author}{\bibfnamefont{A.}~\bibnamefont{Covello}},
  \bibinfo{author}{\bibfnamefont{A.}~\bibnamefont{Gargano}}, \bibnamefont{and}
  \bibinfo{author}{\bibfnamefont{T.~T.~S.} \bibnamefont{Kuo}},
  \bibinfo{journal}{Phys. Rev. C} \textbf{\bibinfo{volume}{68}},
  \bibinfo{pages}{034320} (\bibinfo{year}{2003}).

\bibitem[{\citenamefont{Coraggio
  et~al.}(2005{\natexlab{b}})\citenamefont{Coraggio, Covello, Gargano, Itaco,
  Kuo, and Machleidt}}]{Coraggio05b}
\bibinfo{author}{\bibfnamefont{L.}~\bibnamefont{Coraggio}},
  \bibinfo{author}{\bibfnamefont{A.}~\bibnamefont{Covello}},
  \bibinfo{author}{\bibfnamefont{A.}~\bibnamefont{Gargano}},
  \bibinfo{author}{\bibfnamefont{N.}~\bibnamefont{Itaco}},
  \bibinfo{author}{\bibfnamefont{T.~T.~S.} \bibnamefont{Kuo}},
  \bibnamefont{and}
  \bibinfo{author}{\bibfnamefont{R.}~\bibnamefont{Machleidt}},
  \bibinfo{journal}{Phys. Rev. C} \textbf{\bibinfo{volume}{71}},
  \bibinfo{pages}{014307} (\bibinfo{year}{2005}{\natexlab{b}}).

\bibitem[{\citenamefont{Coraggio
  et~al.}(2006{\natexlab{b}})\citenamefont{Coraggio, Covello, Gargano, Itaco,
  and Kuo}}]{Coraggio06b}
\bibinfo{author}{\bibfnamefont{L.}~\bibnamefont{Coraggio}},
  \bibinfo{author}{\bibfnamefont{A.}~\bibnamefont{Covello}},
  \bibinfo{author}{\bibfnamefont{A.}~\bibnamefont{Gargano}},
  \bibinfo{author}{\bibfnamefont{N.}~\bibnamefont{Itaco}}, \bibnamefont{and}
  \bibinfo{author}{\bibfnamefont{T.~T.~S.} \bibnamefont{Kuo}},
  \bibinfo{journal}{Phys. Rev. C} \textbf{\bibinfo{volume}{73}},
  \bibinfo{pages}{014304} (\bibinfo{year}{2006}{\natexlab{b}}).

\bibitem[{\citenamefont{Kuckei et~al.}(2003)\citenamefont{Kuckei, Montani,
  M{\"u}ther, and Sedrakian}}]{Kuckei03}
\bibinfo{author}{\bibfnamefont{J.}~\bibnamefont{Kuckei}},
  \bibinfo{author}{\bibfnamefont{F.}~\bibnamefont{Montani}},
  \bibinfo{author}{\bibfnamefont{H.}~\bibnamefont{M{\"u}ther}},
  \bibnamefont{and}
  \bibinfo{author}{\bibfnamefont{A.}~\bibnamefont{Sedrakian}},
  \bibinfo{journal}{Nucl. Phys. A} \textbf{\bibinfo{volume}{723}},
  \bibinfo{pages}{32} (\bibinfo{year}{2003}).

\bibitem[{\citenamefont{Sedrakian et~al.}(2003)\citenamefont{Sedrakian, Kuo,
  M{\"u}ther, and Schuck}}]{Sedrakian03}
\bibinfo{author}{\bibfnamefont{A.}~\bibnamefont{Sedrakian}},
  \bibinfo{author}{\bibfnamefont{T.~T.~S.} \bibnamefont{Kuo}},
  \bibinfo{author}{\bibfnamefont{H.}~\bibnamefont{M{\"u}ther}},
  \bibnamefont{and} \bibinfo{author}{\bibfnamefont{P.}~\bibnamefont{Schuck}},
  \bibinfo{journal}{Phys. Lett. B} \textbf{\bibinfo{volume}{576}},
  \bibinfo{pages}{68} (\bibinfo{year}{2003}).

\bibitem[{\citenamefont{Bogner et~al.}(2005)\citenamefont{Bogner, Schwenk,
  Furnstahl, and Nogga}}]{Bogner05}
\bibinfo{author}{\bibfnamefont{S.~K.} \bibnamefont{Bogner}},
  \bibinfo{author}{\bibfnamefont{A.}~\bibnamefont{Schwenk}},
  \bibinfo{author}{\bibfnamefont{R.~J.} \bibnamefont{Furnstahl}},
  \bibnamefont{and} \bibinfo{author}{\bibfnamefont{A.}~\bibnamefont{Nogga}},
  \bibinfo{journal}{Nucl. Phys. A} \textbf{\bibinfo{volume}{763}},
  \bibinfo{pages}{59} (\bibinfo{year}{2005}).

\bibitem[{\citenamefont{Andreozzi}(1996)}]{Andreozzi96}
\bibinfo{author}{\bibfnamefont{F.}~\bibnamefont{Andreozzi}},
  \bibinfo{journal}{Phys. Rev. C} \textbf{\bibinfo{volume}{54}},
  \bibinfo{pages}{684} (\bibinfo{year}{1996}).

\bibitem[{\citenamefont{Suzuki and Lee}(1980)}]{Suzuki80}
\bibinfo{author}{\bibfnamefont{K.}~\bibnamefont{Suzuki}} \bibnamefont{and}
  \bibinfo{author}{\bibfnamefont{S.~Y.} \bibnamefont{Lee}},
  \bibinfo{journal}{Prog. Theor. Phys.} \textbf{\bibinfo{volume}{64}},
  \bibinfo{pages}{2091} (\bibinfo{year}{1980}).

\bibitem[{\citenamefont{Entem and Machleidt}(2002)}]{Entem02}
\bibinfo{author}{\bibfnamefont{D.~R.} \bibnamefont{Entem}} \bibnamefont{and}
  \bibinfo{author}{\bibfnamefont{R.}~\bibnamefont{Machleidt}},
  \bibinfo{journal}{Phys. Rev. C} \textbf{\bibinfo{volume}{66}},
  \bibinfo{pages}{014002} (\bibinfo{year}{2002}).

\bibitem[{\citenamefont{Machleidt and Entem}(2005)}]{Machleidt05}
\bibinfo{author}{\bibfnamefont{R.}~\bibnamefont{Machleidt}} \bibnamefont{and}
  \bibinfo{author}{\bibfnamefont{D.~R.} \bibnamefont{Entem}},
  \bibinfo{journal}{J. Phys. G} \textbf{\bibinfo{volume}{31}},
  \bibinfo{pages}{S1235} (\bibinfo{year}{2005}).

\bibitem[{Ent()}]{Entem07}
\bibinfo{note}{D. R. Entem and R. Machleidt, to be published.}

\bibitem[{\citenamefont{Kuo et~al.}(1981)\citenamefont{Kuo, Shurpin, Tam,
  Osnes, and Ellis}}]{Kuo81}
\bibinfo{author}{\bibfnamefont{T.~T.~S.} \bibnamefont{Kuo}},
  \bibinfo{author}{\bibfnamefont{J.}~\bibnamefont{Shurpin}},
  \bibinfo{author}{\bibfnamefont{K.~C.} \bibnamefont{Tam}},
  \bibinfo{author}{\bibfnamefont{E.}~\bibnamefont{Osnes}}, \bibnamefont{and}
  \bibinfo{author}{\bibfnamefont{P.~J.} \bibnamefont{Ellis}},
  \bibinfo{journal}{Ann. Phys. (NY)} \textbf{\bibinfo{volume}{132}},
  \bibinfo{pages}{237} (\bibinfo{year}{1981}).

\bibitem[{\citenamefont{Shurpin et~al.}(1983)\citenamefont{Shurpin, Kuo, and
  Strottman}}]{Shurpin83}
\bibinfo{author}{\bibfnamefont{J.}~\bibnamefont{Shurpin}},
  \bibinfo{author}{\bibfnamefont{T.~T.~S.} \bibnamefont{Kuo}},
  \bibnamefont{and}
  \bibinfo{author}{\bibfnamefont{D.}~\bibnamefont{Strottman}},
  \bibinfo{journal}{Nucl. Phys. A} \textbf{\bibinfo{volume}{408}},
  \bibinfo{pages}{310} (\bibinfo{year}{1983}).

\bibitem[{\citenamefont{Shurpin et~al.}(1977)\citenamefont{Shurpin, Strottman,
  Kuo, Conze, and Manakos}}]{Shurpin77}
\bibinfo{author}{\bibfnamefont{J.}~\bibnamefont{Shurpin}},
  \bibinfo{author}{\bibfnamefont{D.}~\bibnamefont{Strottman}},
  \bibinfo{author}{\bibfnamefont{T.~T.~S.} \bibnamefont{Kuo}},
  \bibinfo{author}{\bibfnamefont{M.}~\bibnamefont{Conze}}, \bibnamefont{and}
  \bibinfo{author}{\bibfnamefont{P.}~\bibnamefont{Manakos}},
  \bibinfo{journal}{Phys Lett. B} \textbf{\bibinfo{volume}{69}},
  \bibinfo{pages}{395} (\bibinfo{year}{1977}).

\bibitem[{\citenamefont{Kuo and Brown}(1966)}]{Kuo66}
\bibinfo{author}{\bibfnamefont{T.~T.~S.} \bibnamefont{Kuo}} \bibnamefont{and}
  \bibinfo{author}{\bibfnamefont{G.~E.} \bibnamefont{Brown}},
  \bibinfo{journal}{Nucl. Phys.} \textbf{\bibinfo{volume}{85}},
  \bibinfo{pages}{40} (\bibinfo{year}{1966}).

\bibitem[{\citenamefont{Engeland}()}]{EngelandSMC}
\bibinfo{author}{\bibfnamefont{T.}~\bibnamefont{Engeland}}, \bibinfo{note}{the
  Oslo shell-model code 1991-2006, unpublished}.

\bibitem[{\citenamefont{Blomqvist and Molinari}(1968)}]{Blomqvist68}
\bibinfo{author}{\bibfnamefont{J.}~\bibnamefont{Blomqvist}} \bibnamefont{and}
  \bibinfo{author}{\bibfnamefont{A.}~\bibnamefont{Molinari}},
  \bibinfo{journal}{Nucl. Phys. A} \textbf{\bibinfo{volume}{106}},
  \bibinfo{pages}{545} (\bibinfo{year}{1968}).

\bibitem[{\citenamefont{Audi et~al.}(2003)\citenamefont{Audi, Wapstra, and
  Thibault}}]{Audi03}
\bibinfo{author}{\bibfnamefont{G.}~\bibnamefont{Audi}},
  \bibinfo{author}{\bibfnamefont{A.~H.} \bibnamefont{Wapstra}},
  \bibnamefont{and} \bibinfo{author}{\bibfnamefont{C.}~\bibnamefont{Thibault}},
  \bibinfo{journal}{Nucl. Phys. A} \textbf{\bibinfo{volume}{729}},
  \bibinfo{pages}{337} (\bibinfo{year}{2003}).

\bibitem[{\citenamefont{Brown and Richter}(2006)}]{Richter06}
\bibinfo{author}{\bibfnamefont{B.~A.} \bibnamefont{Brown}} \bibnamefont{and}
  \bibinfo{author}{\bibfnamefont{W.~A.} \bibnamefont{Richter}},
  \bibinfo{journal}{Phys. Rev. C} \textbf{\bibinfo{volume}{74}},
  \bibinfo{pages}{034315} (\bibinfo{year}{2006}).

\bibitem[{\citenamefont{Wildenthal}(1984)}]{Wildenthal84}
\bibinfo{author}{\bibfnamefont{B.~H.} \bibnamefont{Wildenthal}},
  \bibinfo{journal}{Prog. Part. Nucl. Phys.} \textbf{\bibinfo{volume}{11}},
  \bibinfo{pages}{5} (\bibinfo{year}{1984}).

\bibitem[{\citenamefont{Brown and Wildenthal}(1988)}]{Brown88}
\bibinfo{author}{\bibfnamefont{B.~A.} \bibnamefont{Brown}} \bibnamefont{and}
  \bibinfo{author}{\bibfnamefont{B.~H.} \bibnamefont{Wildenthal}},
  \bibinfo{journal}{Ann. Rev. Nucl. Part. Sci.} \textbf{\bibinfo{volume}{38}},
  \bibinfo{pages}{29} (\bibinfo{year}{1988}).

\bibitem[{\citenamefont{Fauerbach et~al.}(1996)\citenamefont{Fauerbach,
  Morrissey, Benenson, Brown, Hellstr{\"o}m, Kelley, Kryger, Pfaff, Powell, and
  Sherrill}}]{Fauerbach96}
\bibinfo{author}{\bibfnamefont{M.}~\bibnamefont{Fauerbach}},
  \bibinfo{author}{\bibfnamefont{D.~J.} \bibnamefont{Morrissey}},
  \bibinfo{author}{\bibfnamefont{W.}~\bibnamefont{Benenson}},
  \bibinfo{author}{\bibfnamefont{B.~A.} \bibnamefont{Brown}},
  \bibinfo{author}{\bibfnamefont{M.}~\bibnamefont{Hellstr{\"o}m}},
  \bibinfo{author}{\bibfnamefont{J.~H.} \bibnamefont{Kelley}},
  \bibinfo{author}{\bibfnamefont{R.~A.} \bibnamefont{Kryger}},
  \bibinfo{author}{\bibfnamefont{R.}~\bibnamefont{Pfaff}},
  \bibinfo{author}{\bibfnamefont{C.~F.} \bibnamefont{Powell}},
  \bibnamefont{and} \bibinfo{author}{\bibfnamefont{B.~M.}
  \bibnamefont{Sherrill}}, \bibinfo{journal}{Phys. Rev. C}
  \textbf{\bibinfo{volume}{53}}, \bibinfo{pages}{647} (\bibinfo{year}{1996}).

\bibitem[{\citenamefont{Tarasov et~al.}(1997)\citenamefont{Tarasov, Allatt,
  Ang{\'e}lique, Anne, Borcea, Dlouhy, Donzaud, Gr{\'e}vy, Guillemaud-Mueller,
  Lewitowicz et~al.}}]{Tarasov97}
\bibinfo{author}{\bibfnamefont{O.}~\bibnamefont{Tarasov}},
  \bibinfo{author}{\bibfnamefont{R.}~\bibnamefont{Allatt}},
  \bibinfo{author}{\bibfnamefont{J.~C.} \bibnamefont{Ang{\'e}lique}},
  \bibinfo{author}{\bibfnamefont{R.}~\bibnamefont{Anne}},
  \bibinfo{author}{\bibfnamefont{C.}~\bibnamefont{Borcea}},
  \bibinfo{author}{\bibfnamefont{Z.}~\bibnamefont{Dlouhy}},
  \bibinfo{author}{\bibfnamefont{C.}~\bibnamefont{Donzaud}},
  \bibinfo{author}{\bibfnamefont{S.}~\bibnamefont{Gr{\'e}vy}},
  \bibinfo{author}{\bibfnamefont{D.}~\bibnamefont{Guillemaud-Mueller}},
  \bibinfo{author}{\bibfnamefont{M.}~\bibnamefont{Lewitowicz}},
  \bibnamefont{et~al.}, \bibinfo{journal}{Phys. Lett. B}
  \textbf{\bibinfo{volume}{409}}, \bibinfo{pages}{64} (\bibinfo{year}{1997}).

\bibitem[{\citenamefont{Brown}(2001)}]{Brown01}
\bibinfo{author}{\bibfnamefont{B.~A.} \bibnamefont{Brown}},
  \bibinfo{journal}{Prog. Part. Nucl. Phys.} \textbf{\bibinfo{volume}{47}},
  \bibinfo{pages}{517} (\bibinfo{year}{2001}).

\bibitem[{\citenamefont{Ellis et~al.}(2005)\citenamefont{Ellis, Engeland,
  Hjorth-Jensen, Kartamyshev, and Osnes}}]{Ellis05}
\bibinfo{author}{\bibfnamefont{P.~J.} \bibnamefont{Ellis}},
  \bibinfo{author}{\bibfnamefont{T.}~\bibnamefont{Engeland}},
  \bibinfo{author}{\bibfnamefont{M.}~\bibnamefont{Hjorth-Jensen}},
  \bibinfo{author}{\bibfnamefont{M.~P.} \bibnamefont{Kartamyshev}},
  \bibnamefont{and} \bibinfo{author}{\bibfnamefont{E.}~\bibnamefont{Osnes}},
  \bibinfo{journal}{Phys. Rev. C} \textbf{\bibinfo{volume}{71}},
  \bibinfo{pages}{034401} (\bibinfo{year}{2005}).

\bibitem[{\citenamefont{Zuker}(2003)}]{Zuker03}
\bibinfo{author}{\bibfnamefont{A.~P.} \bibnamefont{Zuker}},
  \bibinfo{journal}{Phys. Rev. Lett.} \textbf{\bibinfo{volume}{90}},
  \bibinfo{pages}{042502} (\bibinfo{year}{2003}).

\bibitem[{\citenamefont{Stoks et~al.}(1993)\citenamefont{Stoks, Klomp,
  Rentmeester, and de~Swart}}]{Nijmegen93}
\bibinfo{author}{\bibfnamefont{V.~G.~J.} \bibnamefont{Stoks}},
  \bibinfo{author}{\bibfnamefont{R.~A.~M.} \bibnamefont{Klomp}},
  \bibinfo{author}{\bibfnamefont{M.~C.~M.} \bibnamefont{Rentmeester}},
  \bibnamefont{and} \bibinfo{author}{\bibfnamefont{J.~J.}
  \bibnamefont{de~Swart}}, \bibinfo{journal}{Phys. Rev. C}
  \textbf{\bibinfo{volume}{48}}, \bibinfo{pages}{792} (\bibinfo{year}{1993}).

\bibitem[{SM9()}]{SM99}
\bibinfo{note}{R.A. Arndt, I.I. Strakowsky, and R.L. Workman, George Washington
  University Data Analysis Center (formerly VPI SAID facility), solution of
  summer 1999 (SM99).}

\bibitem[{nnd()}]{nndc}
\bibinfo{note}{Data extracted using the NNDC On-line Data Service from the
  ENSDF database, file revised as of May 30, 2006.}

\end{thebibliography}

\end{document}